\documentstyle[editedvolume,numreferences]{crckapb} 

\input{amssym.def}
\input{amssym.tex}

\def\caja{\mathsurround=0pt}
\def\eqalign#1{\,\vcenter{\openup1\jot \caja
        \ialign{\strut \hfil$\displaystyle{##}$&$
        \displaystyle{{}##}$\hfil\crcr#1\crcr}}\,}
\def\car{\mathop{\square}}
\def\carre#1#2{\raise 2pt\hbox{$\scriptstyle #1$}\car_{#2}}

\begin{opening}
\title{BPS Saturated Amplitudes and
        Non-Perturbative String Theory.}

\subtitle{Carg{\`e}se 1997, CPTH.PC571.1297, hep-th/9712079}
\author{P.~Vanhove}
\institute{Centre de Physique Th{\'e}orique
           {\'E}cole Polytechnique\\
            91128 Palaiseau, France\\
            vanhove@pth.polytechnique.fr}
\end{opening}

\runningtitle{Non-Perturbative Strings Effects}

\begin{document}
\begin{abstract}
The study of the special $F^4$ and $R^4$ in the effective action for the
Spin(32)/{\bf Z}$_2$ and Type~II strings sheds some light on 
D-brane calculus and on instanton contribution counting. The D-instanton
case is discussed separately.
\end{abstract}
\vskip -.3cm
\noindent
Computations of non-perturbative contributions is still a difficult task.
In (supersymmetric) Field theory a direct computation  requires the
construction of the measure of integration over 
vacua containing $n$-instantons, which is not yet available for every $n$.
The use of dualities is still the main tool for getting the answer.
In String theory the D-instantons contributions for the $R^4$ were conjectured
thanks to the $Sl(2,{\bf Z})$ invariance of the Type~IIb theory
\cite{greengutperle}, and for the
Spin$(32)/{\bf Z}_2$ some {\sl ad hoc\/} rules for the contributions for the
$R^4$ and $F^4$ were derived from the S-duality relation between the Heterotic
and Type~I strings \cite{bfkov}.
In this proceeding I will explain how these non-perturbative contributions
can be computed exactly, concentrating on the eight-dimensional case. Details
and generalizations will be found in \cite{bachasvanhove}.

On the Heterotic side the $F^4$ and $R^4$ terms are extracted from
the fourth point amplitude on the torus involving four vertex operators. In order
to get a non zero contribution, eight fermionic zero modes have to be
soaked up. So all the fermionic zero modes from the
supersymmetric side of the vertex operators have to be saturated, leaving a complete freedom on the non-supersymmetrical current algebra side.
These contributions involve only states belonging to the short multiplets of
the $N_{4d}=4$ supersymmetry algebra \cite{bachaskiritsis}.
The non-perturbative contributions in eight-dimensions  read

\begin{equation}
{\cal L}_{\rm non-pert}= - {2V^{(8)}\over 2^{10}\pi^6} \sum_{N>0} e^{-2i\pi
  N T_s {\cal T}_2} {1\over N}\sum_{N=mn\atop 0\leq j <n} [{\cal D}{\cal
  A}]\left(m\,{\cal U}+j\over n\right)\ .
\label{nonpert}
\end{equation}

\noindent
In this expression ${\cal T}$ and ${\cal U}$ are the moduli of the two-torus of
compactification, $T_s$ is the tension of the soliton and the ${\cal A}(q,F,R)$ is a modular invariant partition function computed in
the Ramond sector expanded at the eight order with respect to the space-time
fermionic zero modes \cite{genus}. For Type~II string for the $R^4$ terms we
don't have anymore freedom for a current algebra, thus ${\cal A}=1$.  
${\cal D}$ is a differential operator whose appearance is related to
non-holomorphicities of the elliptic genus \cite{bfkov}.
Hereafter only the leading part is needed ${\cal D}=1+\dots$\ . 
These fluctuations around the wrapped D1-brane are
described by a SO$(2N)$ Super-Yang-Mills (SYM) theory ($N$ can be an half-integer). Thanks to 
the holomorphicity of ${\cal A}$, (\ref{nonpert}) can be computed at the
infra-red (IR) fixed point of this SYM theory: $g_{\rm SYM}\rightarrow \infty$
with $\alpha' g^2_{\rm SYM}g^2_s = 1$
\cite{matrix}. This limit is a (8,0) supersymmetrical $\sigma$-model with target space the symmetric orbifold 

\begin{equation}
  {\cal M}=({\bf R}^{8})^N/{\cal S}_N\ .
\label{orbifold}
\end{equation}
The Euclidean action in the light-cone gauge

\begin{equation}
\eqalign{
&S_{\rm E} = \int dt\int_0^{2\pi}{d\sigma\over2\pi} {\rm Tr}
\left[\partial {\bf X}^i \left(\bar
 \partial\delta_{ij}+{i{\cal R}_{ij}/2\pi} \right){\bf X}^j + {\bf
 S}_a\partial {\bf S}_a +\right.\hfill\cr
&\hfill\left.+ {}^{\rm T}\lambda_A^{(P)}\left(\bar \partial\delta^{AB}+
      {i{\cal F}^{AB}/2\pi} \right)\lambda_B^{(P)}+
     {}^{\rm T}\lambda_A^{(A)}\left(\bar
       \partial\delta^{AB} + {i{\cal F}^{AB}/2\pi}
     \right)\lambda_B^{(A)} \right]
}
\end{equation}
 describes the coupling between the fluctuations, ${\bf X}^i$, ${\bf S}_a$
 living in
 the adjoint representation of SO$(2N)$ and
 $\lambda_A$ in vectorial representation of SO$(2N)$, with the background fields

\begin{equation}
\eqalign{
({\cal R}_{ij})_{MN} = -{1\over 8} {\cal R}_{ijkl} ({\bf S}^a_0)_{MP} \gamma^{kl}_{ab} ({\bf S}^b_0)_{PN}\cr
({\cal F}^{AB})_{MN} = -{1\over 8} {\cal F}^{AB}_{kl} ({\bf S}^a_0)_{MP} \gamma^{kl}_{ab} ({\bf S}^b_0)_{PN}}
\quad M,N,P \in \{1,\cdots,2N\}\ .
\end{equation}
It is important to remark that the space-time fermionics zero modes ${\bf
  S}^a_0$ enter only through these tensors.
The various twisted sectors of the Hilbert space constructed on ${\cal M}$ are
  labeled by the action of the permutations $\sigma\in {\cal S}_N$ 
  decomposed in disjoint cycles as $\sigma=[1^{N_1},\dots,N^{N_N}]$ with
  $N=\sum_i i N_i$ \cite{matrix}. A sector with boundary conditions twisted by
  $\sigma$ will contain $8\times \sum_i N_i$ fermionic zero modes.  
As only terms with eight fermionic zero modes are needed, we must have
  $\sigma=[N^1]$.
According to the rules of orbifold computation the determinant is given by the
  partition function on a torus computed over the invariant states
  \cite{ginsparg}:

\begin{equation}
Z= {1\over N!} \sum_{\omega=[N^1]}\sum_{\eta\atop\eta\omega=\omega\eta}
\carre{\eta}{\omega}\ ,
\label{Zorbifold}
\end{equation}

$\carre{\eta}{\omega}$ represents the functional integral

$$
\carre{\eta}{\omega}=\int_{\rm Torus}{\cal D}X{\cal D}S^a{\cal D}\lambda\
e^{-S_{\rm E}} $$
calculated for the boundary conditions specified by $\eta$ and $\omega$.
Yielding exactly the determinant of~(\ref{nonpert})

\begin{equation}
Z= {1\over N}\left[{\cal A}(N{\cal U})+
 \sum_{n_1n_2=N\atop n_2\neq 1}\sum_{j=0}^{n_2-1}{\cal
 A}\left(n_1\, {\cal U}+j\over n_2 \right) \right]\ .
\label{partition}
\end{equation}

The Type~II case, is very similar in spirit,
the fluctuations around a vacua containing wrapped D-branes are 
described by a U$(N)$ SYM model with an IR limit described by a (8,8)
supersymmetrical $\sigma$-model (coupled to background for terms involving less
zero modes).
The big difference here is the existence of a U(1) gauge field freedom allowing
't Hooft fluxes \cite{fluxes}. Theses fluxes constraint the twisted states
to appear as bound states of $(p,q)$-strings, $T_s=|p+q\tau|$
 in~(\ref{nonpert}). This gauge field  is a remnant of the invariance
 under volume preserving diffeomorphism of the D-brane's world-volume theory.
 For the
 Spin$(32)/{\bf Z}_2$ case it was projected out by the $\Omega$ parity
 projection.
 As briefly explained in \cite{greenvanhove}, seeing
 the global topology of the solitonic configuration,
 the gauge configuration contributions to the determinant of fluctuations
  are responsible for the appearance of
 the Bessel function $K_1(\cdots)$. In a T-dual language, this is the origin of the fractionalization of the D-instanton charge $N=nm$ \cite{greengutperle,greengutperletwo}.
The D-instanton case is more tricky because the associated topological SYM
model does not have a well defined IR limit. 
 Nevertheless, 
\cite{greengutperlevanhove} shown that these contributions stem from a one-loop
computation involving the super-graviton compactified on a two-torus with
complex structure $\tau=C^{(0)}+ie^{-\phi}$

\begin{equation}
\eqalign{
{\cal L}^{FT}&= \int^\infty_0 {dt\over t}\; {\rm Str}\left(
({\cal R}_{ijkl} M^{ij}{\tilde M}^{kl})^4 e^{-tS} \right)\cr
&=t_{(8)}t_{(8)}{\cal R}^4{\pi^{3/2}\over \Gamma(3/2)} \int^\infty_0 {dt\over
  t}\  t^{3/2}\!\!
\sum_{(l_1,l_2)\neq(0,0)} e^{-\pi t {|l_1+l_2 \tau|^2\over \tau_2}}\ .
}
\end{equation}
$M^{ij}$ (${\tilde M}^{kl}$) is a Lorentz generator in the adjoint
representation of SO(8) containing two left (right) fermionic zero modes
\cite{gsw},
${\cal R}_{ijkl}$ is the curvature.
The super-trace being computed over $1/2$-BPS multiplet of the IIb supergravity
algebra.
In lower dimensions the super-trace computed over the 1/2-BPS momentum
multiplets of the 
U-duality group \cite{greenvanhove}, yields the weight-3/2 Poincar{\'e} series, of course 1/2-BPS states from fluxes multiplets can contribute too \cite{elitzur}.
In ten dimension this expression can be rewritten as 

\begin{eqnarray}
{\cal L}^{FT}&=&2\zeta(3)\tau_2^{3/2}+{2\pi^2\over 3}\tau_2^{-1/2}+\\
&&8\pi\sum_{N\geq 1} \left(e^{2i\pi N\tau}+c.c.\right)
\left[\sum_{N=mn} {1\over m^2}
\right]\left\lbrace\sqrt{N}\left(1+\sum_{k=1}^\infty {(-1)^k\over (4\pi N\tau_2)^k}
  {\Gamma\left(k-{1\over2}\right)\Gamma\left(k+{3\over2}\right)\over\Gamma\left(1\over 2\right)\Gamma\left(3\over 2\right) k!}\right)\right\rbrace\ .\nonumber
\end{eqnarray}
The contribution in square bracket $[\cdots]$ should come from the same
counting of zero modes as before \cite{greengutperletwo}. The contribution in braces
$\lbrace\cdots\rbrace$ is the signature of higher loops corrections around the
D-instantons.


\acknowledgements
I want to thank the ({\sl gentils}) organizers of the Carg{\`e}se summer school for
financial support, and allowing me to give a talk. These works were done in
collaboration with C.~Bachas, C.~Fabre,  M.B.~Green, M.~Gutperle, E.~Kiritsis,
N.A.~Obers. Moreover it is a pleasure to thanks K.~F{\"o}rger, P.~Windey,
F.~Laudenbach and N.~Berline for interesting discussions. 


\end{document}